# Snowpark: Performant, Secure, User-Friendly Data Engineering and AI/ML Next To Your Data

Brandon Baker, Elliott Brossard, Chenwei Xie, Zihao Ye, Deen Liu, Yijun Xie, Arthur Zwiegincew, Nitya Kumar Sharma, Gaurav Jain, Eugene Retunsky, Mike Halcrow, Derek Denny-Brown, Istvan Cseri, Tyler Akidau, Yuxiong He

*Snowflake, Inc*

*Abstract*—Snowflake revolutionized data analytics with an elastic architecture that decouples compute and storage, enabling scalable solutions supporting data architectures like data lake, date warehouse, data lakehouse, and data mesh. Building on this foundation, Snowflake has advanced its AI Data Cloud vision by introducing Snowpark, a managed turnkey solution that supports data engineering and AI/ML workloads using Python and other programming languages.

This paper outlines Snowpark's design objectives towards high performance, strong security and governance, and ease of use. We detail the architecture of Snowpark, highlighting its elastic scalability and seamless integration with Snowflake core compute infrastructure. This includes leveraging Snowflake control plane for distributed computing and employing a secure sandbox for isolating Snowflake SQL workloads from Snowpark executions. Additionally, we present core innovations in Snowpark that drive further performance enhancements, such as query initialization latency reduction through Python package caching, improved workload scheduling for customized workloads, and data skew management via efficient row redistribution. Finally, we showcase real-world case studies that illustrate Snowpark's efficiency and effectiveness for large-scale data engineering and AI/ML tasks.

## I. INTRODUCTION

Snowflake elastic data cloud [1] was introduced as a multi-cloud offering in June 2015. Snowflake is designed as a multi-tenant, transactional, secure, and highly scalable system, with elasticity to handle diverse workloads. The system operates on a consumption model across multiple cloud providers, enabling customers to seamlessly manage and query their data through familiar tools and interfaces. Snowflake's novel multi-cluster, shared-data architecture innovations unblock a wide variety of data analytics use cases for customers. And we see strong adoption momentum on Snowflake's data cloud solution.

As customer needs evolve, there is growing interest to go beyond Snowflake's core data analytics capabilities. A common pattern we observed was that customers would conduct certain data analytics activities in Snowflake, but transferred data out to other systems, such as Spark [14], for data engineering or AI/ML tasks, and moved the results back to Snowflake for further analytics. This approach has multiple drawbacks for customers. The data movements across system boundaries incur high latency, ingress and egress cost, and security risks. It also requires customers to set up and maintain certain additional infrastructure, which is a burden and distraction from handling customers' core business logics. To address these obstacles for customers, we made the decision to introduce Snowpark as the extension to allow customers to run user-authored workloads in Python and other programming languages to process data within Snowflake. We focused on three key objectives as we designed Snowpark:

- High Performance - Customers adopt Snowflake for large volume data analytics. Therefore, Snowpark needs to be price/performance competitive to serve large-scale data processing needs.
- Security and Governance - Customers put business critical data and workloads in Snowflake. Therefore, robust security and governance are essential objectives for Snowpark.
- Ease of Use - We'd like customers to focus on their own business logic rather than worrying about tuning the underlying infrastructure.

To meet the above objectives, Snowpark builds on top of Snowflake's virtual warehouse compute model and manages secure sandboxes in virtual warehouses to isolate SQL data analytics and extended compute. This architecture simplifies the compute infrastructure management experience for customers, while guaranteeing strong security enforcement. Snowpark leverages Snowflake's control plane to distribute compute workloads across workers in virtual warehouses, sharing the scalability and elasticity of Snowflake. The shared control plane provides the foundation for frictionless integration between Snowpark and Snowflake's SQL compute, eliminating the needs of moving data outside of Snowflake governance for extended processing.

This paper presents the Snowpark architecture, detailing its essential components. Recognizing Python's significant impact in data engineering and AI/ML areas, we present the Snowpark Python interface, which encompasses stored procedures, the DataFrame API, and user defined functions (UDFs). We discuss the execution model for Python functions and illustrate how it leverages Snowflake's existing infrastructure for scalability and ease of use. Furthermore, we explain the sandbox design and outline the security mechanisms for Snowpark's stringent security requirements. Next, we present several critical innovations about performance optimization on top of the core architecture. And lastly, we conduct real production case studies on common use cases to show how Snowpark brings performance, security, and ease of use advantages to customers.

The rest of this paper is organized as follows. Section 2 covers the background of Snowflake architecture. Section 3 presents the Snowpark architecture built on top of existing Snowflake infrastructure. We then discuss key performance optimizations in Section 4 and show real-world production case studies in Section 5. Finally, Section 6 concludes the paper.

## II. BACKGROUND - SNOWFLAKE OVERVIEW

Snowflake [1] is built to be an enterprise-ready service, prioritizing high usability, interoperability, and availability. It utilizes a service-oriented architecture that is both highly fault-tolerant and independently scalable. Services are organized into three main architectural layers.

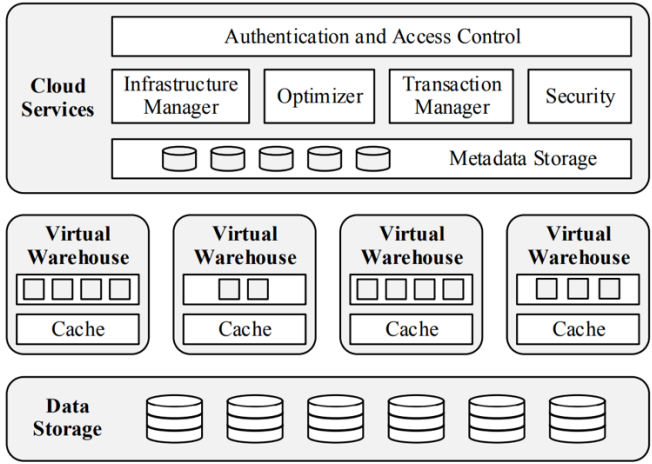

Fig. 1. Snowflake Architecture

- Data Storage - This layer leverages cloud providers' blob storages (e.g. Amazon S3 [3], Azure Blob Storage [4], Google Cloud Storage [5]) to store both table data and query results efficiently.
- Virtual Warehouses: Serving as the "muscle" of the system, this layer runs queries using elastic clusters of virtual machines managed by Snowflake.
- Cloud Services: Acting as the "brain", which consists of a suite of services that manage virtual warehouses, queries, transactions, as well as various metadata, such as database schemas, access control details, encryption keys, and usage statistics.

Snowflake's architecture provides several advantages that make Snowflake highly adaptable to a wide range of use cases. The separation of storage and compute allows for independent scaling, meaning that resources can be adjusted based on the specific needs of different workloads without over-provisioning. This results in cost efficiency and performance optimization, making Snowflake suitable for both small scale analytics and large scale data processing. The use of cloud services to manage transactions, metadata, and security ensures robust performance, high availability, and strong governance.

## III. SNOWPARK ARCHITECTURE

Snowpark is a combination of libraries and code execution environments that enable customers to run workloads in Python and other programming languages next to their data in Snowflake. Snowpark provides a DataFrame API to accept well-defined DataFrame operations and emits corresponding SQL queries to execute in Snowflake, fully leveraging the scalability and elasticity of Snowflake's SQL query engine. Furthermore, Snowpark provides the code execution environments where customers' arbitrary user code can be pushed down to execute in Snowpark's secure sandbox, which is the infrastructure to isolate customers' Snowpark workloads from Snowflake SQL workloads. To minimize performance overhead, security risk, and simplify customers' resource management experience, Snowpark fits the computation into Snowflake's virtual warehouse model, where Snowpark secure sandboxes are provisioned in Snowflake virtual warehouses for user code execution, and share the same virtual warehouse compute resources.

Given the popularity and significant impact of Python, in this section we will go through the key Snowpark components for Python and how they interact with each other, as well as Snowflake to achieve the Performance, Security, and Ease of Use objectives.

### A. Snowpark Python Interfaces

Snowpark enables users to run Python programs as Python stored procedures. Within stored procedures, users can run arbitrary Python code, including issuing queries to Snowflake. Snowpark builds a Python DataFrame API to allow developers to write data processing logic directly in Python. The API layer takes Python DataFrame operations, and emits corresponding SQL statements to execute in Snowflake.

Snowpark also supports User Defined Functions in Python and other languages to process data in Snowflake. A UDF is executed on a per row basis, and can be invoked as part of SQL statements. Snowflake handles data type conversions and data transmission between the SQL layer and Python layer.

Vectorized processing has been a very common approach [6, 7] to boost performance for popular Python packages, such as NumPy [8], TensorFlow [9], PyTorch [10], etc. With the objective of fully leveraging such performance boost in Python packages, we introduced vectorized interfaces for Python UDFs, where users can annotate UDFs to change the Python UDF from row-based processing to DataFrame processing. We adopted Pandas [2] DataFrame as the concrete implementation.

In addition to user defined scalar functions, Snowpark also supports User Defined Table Functions (UDTFs) and User Defined Aggregate Functions (UDAFs). UDTFs return a set of rows (i.e. a table), and UDAFs operate on multiple rows of data and return a single aggregated result per group.

### B. Python Function Execution

Python Functions are executed as part of Snowflake SQL queries. At query startup time, depending on users' Python code, necessary Python packages and imports are installed on the Snowflake virtual warehouse nodes within the Snowpark sandbox. Since Python prior to 3.13 has a global interpreter lock, Snowpark creates many Python interpreter processes for each function in the query. Snowpark initializes the Python interpreter before forking additional processes to reduce initialization time. The virtual warehouse worker threads communicate with the Snowpark Python interpreter processes through gRPC [11] to pass rowsets [12] for computation. Figure 2 presents the compute model for Snowpark Python functions and how it fits into Snowflake architecture.

When the query is done, Python packages and imports remain in a local cache but the sandbox and Python interpreters are cleaned up. Re-running the same query, or another query that uses some of the same packages or imports, will be faster due to the cache that Snowpark maintains.

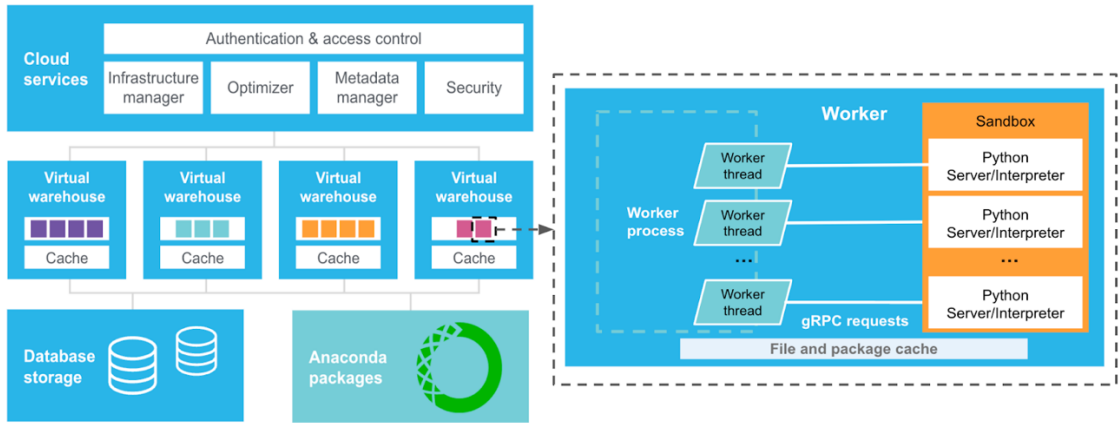

Fig. 2. Snowpark Python Function Compute Model

*C. Secure Sandbox*

Unlike SQL workloads which run defined SQL statements in a constrained language, Snowpark runs arbitrary user code inside of our virtual warehouse environment. As a result, it is at greater risk of exploitation. Therefore, Snowpark adopts a secure sandbox comprised of several layers of defense as well as additional security controls to achieve its security objectives of strong isolation, data governance, and access control.

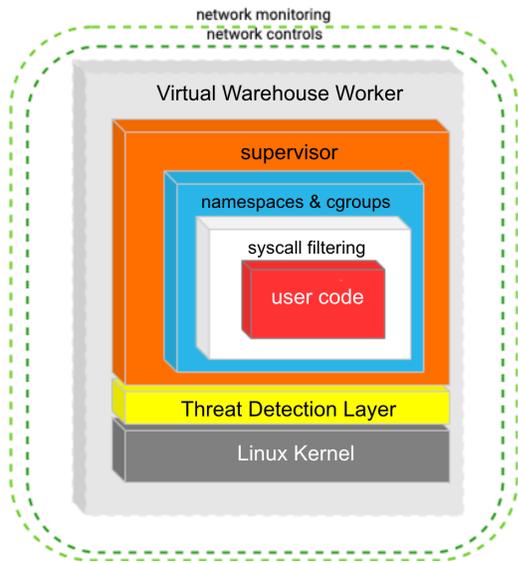

Fig. 3. Snowpark Sandbox

- Namespaces and cgroups: We use namespaces to isolate processes, and cgroups to manage resources, such as CPU and memory. This isolation allows different workloads to run independently without interfering with each other.
- Syscall filtering: We introduced a syscall filtering layer to make sure insecure syscalls are blocked. The layer maintains a list of allowed or conditionally allowed syscalls and denies other potentially malicious syscalls.
- Supervisor Process: We introduced a supervisor process to provide the logging capabilities to track all denied syscalls in the sandbox. We leverage these logging data to monitor workloads' patterns and identify potential malicious actors.

Over time, these syscall mechanisms have evolved, which include changes of underlying syscall filtering implementation itself, without appreciable user-visible changes to the code execution environment inside the sandbox. This flexibility has allowed us to improve security over time while also providing more functionality inside the sandbox – for example, adding external network access or support for native code.

In addition to Snowpark's secure sandbox, to mitigate the risk of data exfiltration, we introduce another layer of defense through network egress policies and external egress proxies. These policies are generated by the control plane and enforced at the network edge. This ensures that in the event of compromise of user code inside a sandbox, or even in the event of a sandbox compromise itself, user data cannot be exfiltrated to external destinations outside of predefined, user-specified egress policies.

*D. Architecture Advantages*

Snowpark's architecture integrates with the Snowflake core infrastructure seamlessly, and eliminates the need for users to transfer data outside of Snowflake for extended data processing. This sets the foundation to achieve Snowpark's design goals.

**Performance**: Local data processing avoids the network overhead and data movement cost across systems' boundaries. This architecture also allows Snowpark to inherit Snowflake's scalability and elasticity naturally.

**Security and Governance**: By embedding the extended compute in Snowflake virtual warehouses, Snowpark makes sure the data does not leave Snowflake, and customers can naturally leverage Snowflake's data governance features.

**Ease of Use**: Snowpark's design choice of building on top of Snowflake's existing infrastructure brings the natural benefits that customers don't need to maintain any additional compute resources. All Snowflake's built-in easy-to-use features are also immediately accessible to Snowpark customers.

On top of the Snowpark architecture, we introduce several additional Snowpark infrastructure performance optimizations, which are discussed in the next section.

IV. OPTIMIZATIONS TOWARDS HIGH PERFORMANCE

To address complex data engineering and AI/ML use cases, other than inheriting native Snowflake advantages of scalability,

elasticity and simplicity, this section describes the additional key innovations we've made towards high performance computing.

*A. Python Package Caching*

Caching is a common mechanism in OLAP systems to speed up data processing. Unlike SQL query computation, where data processing starts immediately when data is ready, Snowpark needs to address the additional challenges of efficiently setting up environments for Python code execution, which involves downloading and installing Python packages referenced in users' Python code. To achieve the performance objective, we introduced multiple layers of caching optimizations around Python packages.

**Solver Cache**: To make sure Python code referenced packages work properly in the Sandbox, Snowpark invokes the conda solver to identify the package dependencies. This process is time consuming, especially when users' Python code references multiple packages, where the solver needs to identify the transitive closure of required packages and guarantee that there are no version conflicts among those packages. Snowpark keeps a global solver cache to map package combinations to their corresponding fully expanded package dependencies. When there is a cache hit, this allows Snowpark queries to skip the package dependency solving phase and directly start individual package preparation. Since the cache is around package metadata and global across all customer accounts and virtual warehouses, the solver cache hit rate in production is as high as 99.95%.

**Environment Cache**: Snowpark enables installed package reuse across queries within the same virtual warehouse through the environment cache. When a query finishes in the virtual warehouse, the environment cache will maintain two mappings. One mapping is between the query's package combination and the runtime environment, and the other is between each individual package ID and the actual installed package binary. If a new query lands on the same virtual warehouse, and uses the exact same list of packages as in a previous query, Snowpark will directly load the corresponding runtime environment to start the user code execution. Otherwise, Snowpark will go through new query's package list and create a new runtime environment, leveraging cached individual packages in the virtual warehouse as well as downloading missing packages from the central package repository. The packages are evicted on LRU basis, and the environment cache gets reset when the virtual warehouse machines are recycled by cloud providers. In production, we observe on average 92.58% environment cache hit rate.

Figure 4 illustrates Snowpark Python queries' initialization latency comparison of the three settings against the production workload at P75 (75th percentile), P90, and P95. In general, the solver cache can reduce query initialization latency by around 85%, while the environment cache provides an additional reduction of 65% to 85%. With both caching layers combined, query initialization is accelerated by a factor of 18 to 48, across different percentiles.

In addition to the two caching layers, to speed up cold start latency, as part of the provisioning process, Snowpark will pre-create the root directory to contain necessary system libraries and dependencies as the base environment for Python runtime initialization. Furthermore, we built a Python package prefetch mechanism that prefetches popular Python packages to the virtual warehouse nodes before the first workload starts. Both mechanisms play the role of warming up virtual warehouses for Snowpark Python workloads to reduce cold start latencies.

*B. Historical Stats based Scheduling*

In general, scheduling involves generating estimations for workloads' resource consumption, and placing the workloads in the system correspondingly. While Snowpark naturally inherits workloads placement strategy from Snowflake, it's infeasible to directly adopt Snowflake resource consumption estimations. In classic SQL data warehouses [15, 17], the data processing is broken down into well-defined SQL operators. And since these operators are implemented natively in the query engine, along with the metadata inputs, such as number of rows processed, size of rows, etc, the scheduler is able to calculate the resource consumption. On the other hand, Snowpark workloads run arbitrary user code, making it infeasible to inspect the detailed operations in user code to project the resource consumption.

Instead of adopting a single default value for static resource allocation, or following the other popular user code execution systems, such as Spark [14] or Kubernetes [16], which rely on users to manually annotate the consumption, we introduced an approach to estimate resource consumption based on historical execution stats.

Memory is the primary resource in terms of Snowpark's scheduling consideration, since oversubscribing memory can cause Out Of Memory (OOM) issues and crash workloads. Snowpark built a historical workload execution stats tracking framework. During Snowpark query execution, the query periodically reports the current memory consumption. The framework tracks the max memory consumption through the life cycle of a query and stores that max value in the query's metadata as the current execution's memory consumption stat. When a new execution of the same query starts, it looks back at the past K executions' memory consumption stats, and takes the P percentile value, with a multiplier factor F, as the query's memory consumption estimation.

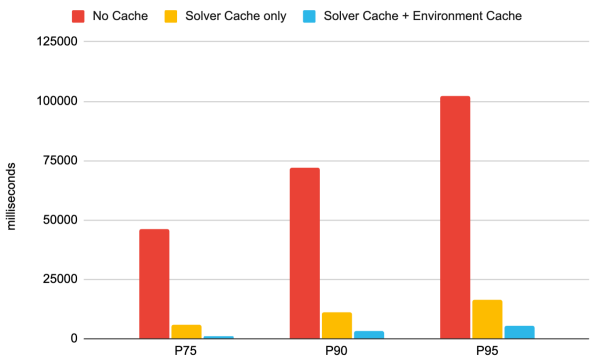

Fig. 4. Query Initialization Latency Comparison

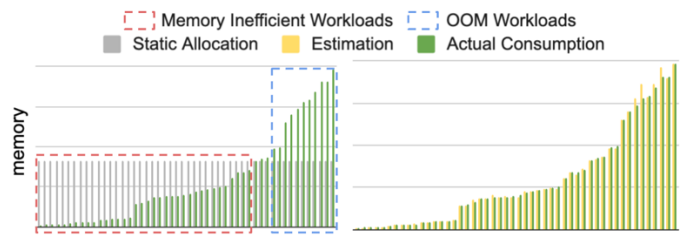

Fig. 5. Static Memory Allocation vs. Dynamic Estimation

Figure 5 presents the visualization of 50 sampled production workloads across different memory consumption ranges to demonstrate the advantage of this approach against the static-memory-allocation approach. While the baseline approach will cause memory wasting for certain workloads, reflected as longer workloads queuing time, and also cause OOM crashes for other workloads, the dynamic-estimation approach can effectively estimate the consumptions, especially for production workloads, which are usually stable, or evolve gradually. This approach helps Snowpark reduce the OOM rate below 0.0005%, while keeping the 90 percentile query queueing time below 5ms in production.

*C. Row Redistribution for User Defined Functions*

Snowflake achieves scalability by distributing workloads to multiple workers within a virtual warehouse. Snowpark also runs multiple Python interpreter processes for highly parallel UDF computing. This architecture setup can introduce data skew issues when unevenly distributed workloads are assigned to worker processes, and lead to performance degradation.

Since Snowpark's Python user code may take a longer time to process a single row, data skew issue can bring more obvious performance degradation to Snowpark, making it a blocker for performant computing. To resolve this issue, we introduced an optimization to redistribute the rows when Snowpark workloads are involved. During the execution stage, the source rowset operator will redistribute the rows across all Python interpreter processes in different virtual warehouse nodes using a round-robin approach, ensuring full parallelism.

Redistribution is not free, and to achieve even distribution, the source rowset operator may need to send rows to the Python interpreter processes in a remote node. This will increase the number of networking calls issued to the processes as well as the overhead to move data between different nodes. When overhead exceeds the impact of data skew, performance is even worse with redistribution applied.

In order to minimize the overhead for redistribution, we examine the workload's per-row execution time from historical stats and define a threshold (T) to determine whether it is worth row level redistribution. Furthermore, to reduce the networking calls for redistributing rows, as the source rowset operator generating output rows, we buffer the rows and asynchronously redistribute them to the target rowset operator when the receiver finishes the previous batch of work.

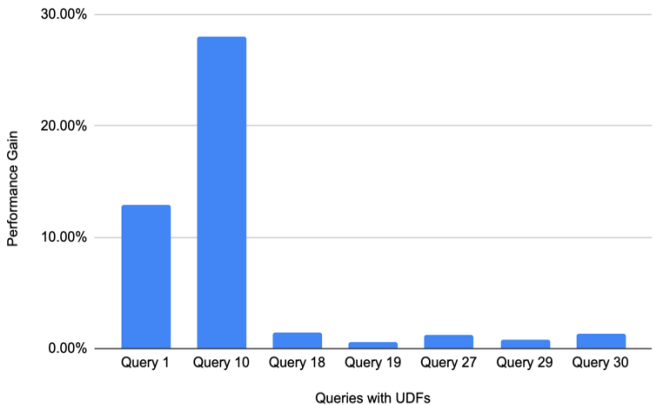

Fig. 6. Performance Gain from Row Redistribution

We validated the redistribution approach using the TPCx-BB [13] benchmark. Among queries with UDFs, we see performance improvements from 0.6% to 28.1%, as illustrated in Figure 6. This matches the above mentioned intuition that not all queries can benefit from row level redistribution, such as cases where rows' distribution is already balanced, or redistribution's overhead offsets the performance gains. In production, we observe redistribution is applied to 37.6% Snowpark UDF queries, and by A/B replaying applicable queries, it shows on average 20.4% performance gain when redistribution is applied.

## V. PRODUCTION CASE STUDIES

Since Snowpark for Python became generally available in June 2022, adoption has grown rapidly. As of June 2024, over 50% of Snowflake customers are using Snowpark, with more than 100 million Snowpark queries running every day. The large number of Snowpark use cases can be categorized into data engineering scenarios and AI/ML scenarios. In below sections, we present case studies in both scenarios from Snowpark customer stories to show the efficiency and effectiveness of Snowpark.

*A. Data Engineering Case Study*

Chicago Trading Company (a.k.a CTC) [18], a leading derivatives trading firm, runs tens of thousands of ETL jobs every day to process various datasets, including feeds from every exchange they trade on, historical trading prices and third-party data. These data engineering jobs transform a large volume of data every night to produce the insights that traders need when markets open the following morning.

Historically, CTC operated Spark [14] clusters for data engineering tasks, and struggled with performance as well as frequent job failures, impacting critical SLAs. With migration from Spark to Snowpark, CTC observed the below advantages:

- Reduced costs and stronger security by eliminating data movement: By moving toward a fully integrated solution, CTC cut down costs by 54%, saving millions each year by getting better visibility into spend and avoiding costly data transfers, which also pose security risks.
- Increased reliability and speed for more efficient data processing: In addition to resolving the reliability issues it saw on managed Spark, the CTC team was able to hit their SLA deadline every day for the first time in the company's history — a milestone it hadn't been in a position to track until Snowpark put the goal within reach.
- Simplified ecosystem for improved usability and innovation: The new system is less taxing on CTC's engineers, who can now report on the ROI of their efforts and focus more on bringing new value to the firm through technological innovation.

Overall, by moving to Snowpark, CTC's research platform brings development to the data for quick, cost-effective and reliable data processing.

*B. AI/ML Case Study*

Fidelity [19] has consolidated company's analytics data into their Enterprise Analytics Platform, which is engineered using

Snowflake. This platform is built with core guiding principles on minimizing data transfers across networks, reducing data duplication, and prioritizing computation directly at the data source whenever possible. Snowpark was chosen for its user-friendly interface and seamless integration with Snowflake, which aligns with Fidelity's guiding principles and translates into cost and performance benefits for customers.

Specifically, Fidelity adopted Snowpark as the key solution to accelerate and simplify feature engineering processes, and observed significant performance improvements across three key scenarios:

- Min-Max Scaling [20]: A critical preprocessing step for preparing data for modeling, where numerical values are scaled into a fixed range between 0 and 1. While the original baseline solution doesn't scale on large datasets, Snowpark eliminates all data movement and brings 77x performance acceleration.
- One-Hot Encoding [21]: A key feature transformation technique for categorical values. Compared to the original baseline solution, Snowpark parallelizes processing for the data transformation and eliminates the data read and write times to bring 50x performance improvements for the customer.
- Pearson Correlation [22]: A statistical measure used to quantify the linear relationship between two variables. By leveraging Snowpark for computing the values, the customer observes 17x performance gain over the original baseline solution.

Overall, Fidelity has achieved significant time, performance and cost benefits by bringing the compute closer to the data. This also has helped to consolidate the compute infrastructure to get rid of unnecessary maintenance overhead.

## VI. CONCLUSION

In summary, Snowpark provides capabilities for users to run Python and other programming language workloads in Snowflake through user code conversion to SQL statements, as well as code execution in Snowpark secure sandboxes, which reside in Snowflake virtual warehouses. By building Snowpark inside Snowflake, the two integrate seamlessly to bring extended compute to customers' data for high performance, strong security and governance, and ease of use. Through additional optimizations, such as package caching mechanisms, historical stats-based scheduling, and data skew handling through row level redistribution, Snowpark brings further performance benefits to diverse workloads, varying from data engineering to AI/ML tasks. These foundational focuses of performance, security, and usability make Snowpark a powerful platform for modern data-driven organizations.

## ACKNOWLEDGMENT

Snowpark is the result of the hard work and dedication of numerous talented individuals, far too many to mention individually. We extend our gratitude to the entire Snowpark team for their remarkable contributions, effort, and passion in developing the product. Our thanks also go out to all the partner teams for their incredible support for Snowpark, and efforts bringing this innovative solution to users. It is an honor and a privilege to work with such an outstanding team, and we are constantly inspired by their excellence.